# Teaching Programming Concepts by Developing Games

Kailash Chandra
Engineering Technology, Pittsburg State University
kchandra@Pittstate.edu

&

Shyamal Suhana Chandra
shyamalc@gmail.com

## Abstract

Learning to program could possibly be analogous to acquiring expertise in abstract mathematics, which may be boring or dull for a majority of students. Thus, among the countless options to approach learning coding [1-14], acquiring concepts through game creation could possibly be the most enriching experience for students. Consequently, it is important to select a lucid and familiar game for students. Then, the following step is to choose a language that introduces the basic concepts of object-oriented programming really well. For this paper, we chose the game of Tic-Tac-Toe, which is straight-forward for most people. The programming language chosen here is C++.

## 1 Introduction

At the introductory level, there are a wide variety of factors [14] influencing the acquisition of programming as a skill by students where meaningful and relevant assignments is one. Studying the programming concepts is more enjoyable when the outcome of the assignment is an interactive game. Also, it is important that the students are already very familiar with the simplicity of the game; thus, games like chess would be highly not recommended. For this paper, we profile the design of Tic-Tac-Toe with C++. The specific approach outlined is supported by [15, 16].

## 2 Tic-Tac-Toe Game

Once more, the game chosen for this paper is Tic-Tac-Toe. In this game, there is a board with nine slots and two players where each make moves alternatively by putting an x or an o, if the move by a player makes a straight set (horizontal, vertical, or diagonal) of three x's or three o's, then that player is declared the winner. If none of the horizontal, vertical, or diagonal slots are filled and there aren't anymore slots to fill, then the game ends in a draw. The following pictures show the progression of a game from start to finish, started by a player and winner with the symbol x:

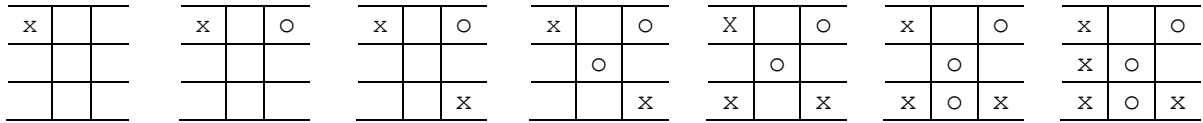

For brevity and simplicity, we assume that in this implementation, a human plays against the computer. There are several variations [17] of this board game. The game can be played human to human, computer to human, human to computer, and computer to computer. Following the diagram below, the design would be to make it possible to go to the previous states of the current game. There will be statistics visible about the games played, including win and draw counts. Other details and controls that are available are shown by the following diagram:

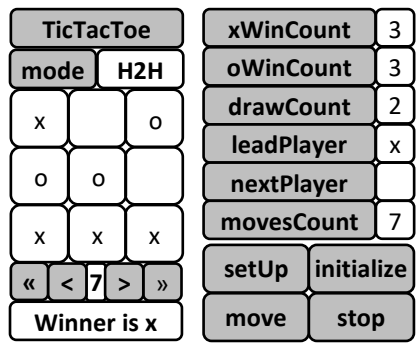

For planning purposes, the diagram shown above with two columns of elements will be used. We will describe all the elements on this diagram with a Tic-tac-toe board and the statistics along with other elements during gameplay:

The first element in the left column of the elements is a label element displaying the name of the game Tic-tac-toe. The second pair of elements shows the **mode** of the game. It could be **H2H** for a human to human game when alternative play between two humans, **H2C** when a human (lead player) is playing against the computer, **C2H** when the computer (lead player) is playing against a human, and **C2C** when the computer is playing against itself. This element can be set by the user by clicking on the **setUp** element. Next, the group element of nine cells shows the tic-tac-toe **board**, where the player can click and make the move. Next, the group of five elements shows the controls to go to the initial ("«") state, previous ("<") state, the move number, control to go to the next (">") state, and the final ("»") state in the current game. The two elements ("«", "»") move the state of the board to the beginning and the last states respectively without changing the statistics of the game set. Next element shows the status or the result of the recent move (e.g., showing Winner is x).

In the right column of elements, **swingout** shows the number of games won by the player x so far, **oWinCount** shows the games won by the player o, **drawCount** shows the number of draws, **leadPlayer** shows the player who made the first move, **nextPlayer** shows the player who is going to make the next move, and **movesCount** shows the number of moves made so far in the current game. The element **setUp** invokes setting up the game parameters, such as the game **mode** (H2H, H2C, C2H, or C2C), the **leadPlayer** (x or o), and other parameters which could be added later to enhance the game. The element **initialize** initializes the board elements to play the next game, the element **move** instructs the computer to make the next move (if the mode is C2C), and element **stop** halts the game.

Initially, these labels convey future actions or values represented by these elements for the developers. Once the software is fully-developed, these labels can be altered so that they are clearly understood by the software's player or user.

The purpose of this diagram is to clearly specify the output required and give an insight into the software's function. This will also allow us to define a class with appropriate data and function members. We will not attempt to address these issues related to the user interface in this paper.

## 3 C++ Implementation

The language we chose for this implementation is C++ [18], which clearly introduces the object-oriented concepts to the students in a very systematic and explicit way. As evident, we will define a class to describe the data and function members associated with the class as follows:

```cpp
class CTicTacToe
{
private:
    string mode;                        // mode of the game H2H, H2C, C2H, or C2C
    char board[3][3];                   // board
    char result;                        // result of the game 'c', 'x', 'o', or 'd'

    int  xWinCount;                     // win count for 'X'
    int  oWinCount;                     // win count for 'O'
    int  drawCount;                     // draw count
    char leadPlayer;                    // first player 'o' or 'x' for all games
    char nextPlayer;                    // next player 'o' or 'x', and blank (' ') if game over
    int  movesCount;                    // moves made so far in current game
    int  nextR;                         // current or next move's row
    int  nextC;                         // current or next move's column
    string history[9];                  // for each move x|o, r, c, and x (9 tuples max for a game)
    int  cursor;                        // the position of move if viewing history of moves
public:
    CTicTacToe(void);                   // default constructor for a game
    ~CticTacToe();                      // destructor for class CTicTacToe
    void setUp(void);                   // sets the game parameters
    void displayCurrentState(void);     // display board with stat
    void getNextMove(void);             // get nextR and nextC for the next player
    void makeNextMove(void);            // make next move based on nextR and nextC
    void checkResult(void);             // sets result to x won | o won | d draw | c continue
    void stop(void);                    // stops current game set and display stat

private:
    void updateResult(void);            // display message after current game is over
    void initialize(void);              // resets board to blanks
    void moveToPreviousState(void);     // update board elements with previous board state
    void moveToNextState(void);         // update board elements with next board state
    void moveToFirstState(void);        // update board elements with first board state
```

```
    void moveToLastState(void);          // update board elements with last board state
};
```

## Main algorithm for the game

```
Create a gameSet of type CTicTacToe
Display gameBoard (3x3 board and the stats)
Do forever
  {
  Get action
  If exit from game desired then exit loop
  Take action
  Update gameBoard
  CheckStatus
  Display Status
  If current game is over
    Initialize gameBoard for new game
    Update gameBoard
    Endif
  }
```

### 3.1 Data Members

The following are the private data members of CTicTacToe class and their brief descriptions:

**string mode**

This is a private data member of the class used to store the **mode** of the game, which can be "H2H", "H2C", "C2H", or "C2C". Alphabet stands for human and alphabet C stands for computer. As a result, "H2H" means that the game is between two humans. Initially it is set to a string "H2H" which could be changed by choosing the setup option for the game. In other words, it could be modified by the public function member **setup**.

**char board[3][3]**

This 3x3 array of characters representing the tic-tac-toe board is used to store the moves made. Initially, all the places in this array are set to blank characters by the default constructor or the function **initialize**. Whenever a player makes a move, an **o** or an **x** is placed in the desired location if that location is available on the board. Access to the board elements is available through **getNextMove** and **makeNextMove** in addition to **initialize** function members.

**char result;**

This private data member is used to store the result of an action, it could be 'c', 'x', 'o', or 'd'. Initially it is set to 'c' meaning game is currently being played.

**int xWinCount;**

This private data member stores the number of wins by the player who plays with 'x' symbol.

**int oWinCount;**

This private data member stores the number of wins by the player who plays with 'o' symbol.

**int drawCount;**

This private data member stores the number of draws by the players in all the games played since the start of the program.

**char leadPlayer;**

This private data member stores the character for the first player, the player that makes the first move for a game, it can be 'o' or 'x'. This will remain true for all the games played until the program is stopped by selecting the **stop** option.

**char nextPlayer;**

This data member stores the symbol for the next player, which could be 'o' or 'x' or ' ' if the current game is over.

**int movesCount;**

This private data member represents the number of moves already made for the current game being played. Initially, this member is set to 0.

**int nextR;**

This private data member will store the next move's row number (0, 1, or 2).

**int nextC;**

This private data member will store the next move's column number (0, 1, or 2).

**string history[9]**

This is a private data member, an array of nine character strings to store the player (x or o), r, and c for the current game (a maximum of 9 tuples for a game). For example, it could store "x00", "o01", "x02", etc. The information in this array will be used by the user to glance through the previous states of the game.

**int cursor;**

This is a private data member used to store the position of state/move in the history array currently being viewed.

### 3.2 Function Members

Following are the public function members of CTicTacToe class:

**CTicTacToe(void)**

The default constructor for the game board that does not take any parameters. It will initialize all the data members to their proper default initial values in order to start the game.

**CTicTacToe(char ch);**

Parameterized constructor for the game board that will initialize all the data members to their proper random values in order to show a random and valid state of the game. Main purpose of this constructor is testing all the other functions of the class.

**~CTicTacToe();**

This is destructor for class CTicTacToe called automatically to release all the allocated memory by an object when the object goes out of scope. It is the programmer's responsibility to properly release all the dynamically allocated memory to avoid memory leak.

**void setup(void);**

This is a public member function used to set some of the game parameters of the game, such as: **mode** ("H2H", "H2C", "C2H", or "C2C") of the game, leading player ("o" or "x") of the game, etc.

**void initialize(void);**

This is a private member function called by a constructor to initialize all the data members (board elements and statistical data points) to properly initialize values in order to start the game.

**void displayCurrentState(void);**

This is a public function to display the 3x3 board and all the statistical (except history) values.

**void getNextMove(void);**

A public member function that determine the row and column number for the next player. It is manually specified or automatically determined depending on the next player. In determining the next player, the value of the game parameter mode ("H2H", "H2C", "C2H", or "C2C") and the current player are used.

**void makeNextMove(void);**

This is a public member function that tries to make the next move based on the information received from getNextMove function, by changing the values for the game board and the stat.

**void checkResult(void);**

A public member function that based on the latest move made, determines if the game should continue further by setting the value of the private data member **result** to 'c' if the current player has made a winning move by returning 'x' or 'o' or if there is a draw by returning 'd'.

**void updateResult(void);**

This is a private function that updates an appropriate message on the status element based on the value of the data member **result** set by the **checkResult** function and updates the stat for the game set. For example, if the after current move, game is over, the display "x Won", "o Won", or "Draw".

**stop(void);**

Public member function, it ends the current game set and exits the program.

**void moveToPreviousState(void);**

This is a private function that fills the board (3x3) with the values representing the previous state based on the current state of the board without changing the stats of the games. It uses the tuples stored in the history array. It also updates the value of **cursor** data member.

**void moveToNextState(void);**

This is a private function that fills the board (3x3) with the values representing the next state based on the current state of the board without changing the stats of the games. It uses the tuples stored in the history array. It also updates the value of **cursor** data member.

**void moveToFirstState(void);**

This is a private function that fills the board (3x3) with the values representing the first state or the beginning state (all values blanked in the 3x3 board) without changing the stats of the games. It uses the tuples stored in the history array. It also updates the value of **cursor** data member (a.k.a. it sets it to zero).

**void moveToNextState(void);**

This is a private function that fills the board (3x3) with the values representing the last state of the board without changing the stats of the games. It uses the tuples stored in the history array. It also updates the value of **cursor** data member (in other words sets it to **movesCount**).

## 4 Concepts Covered

**Class Definition:**

The first concept presented here the concept of a class. It introduces that to solve even a simple problem, it is helpful to group the values and related functionalities and create objects with those values and functionalities. It is similar to the real-world objects. For example, we say that there are three rechargeable eclectic shavers of type S. Each electric shaver has some data values such as: weight, cost, dimensions, etc. and comes with some useful functionalities. It means that S represents a class of shavers. Here we have defined a class named CTicTacToe. If one has to create an object of this class type, it can be created by the following statement:

```
CTicTacToe game;
```

This will create an object named game with its own copy of data members and access to the functions declared in the class definition.

**Data Members:**

In this example, all our data members are private. Each object has its own copy of these values. They are declared private so that cannot be directly altered. Only class function members can alter their values. It is possible to have the data members as public if useful.

**Function Members:**

Each object of CTicTacToe class type will have access to the functions declared in the class. Public function members can be called from the main program, while private function members can only be called by the member functions.

## 5 Conclusion

We believe it is a very simple and interesting way to introduce programming concepts in an object-oriented way. The concepts learned in this development can be easily used in much more complex projects.

**References**

[1] Top 10 Ways to Teach Yourself to Code
https://lifehacker.com/top-10-ways-to-teach-yourself-to-code-1684250889

[2] 27 ways to learn how to code on the cheap (or free)
https://thenextweb.com/dd/2017/04/03/so-you-want-to-be-a-programmer-huh-heres-25-ways-to-learn-online/

[3] 49 of The Best Places to Learn to Code For Free
https://learntocodewith.me/posts/code-for-free/

[4] 12 Sites That Will Teach You Coding for Free
https://www.entrepreneur.com/article/250323

[5] 12 Free Games to Learn Programming
https://medium.mybridge.co/12-free-resources-learn-to-code-while-playing-games-f7333043de11

[6] 10 Great Websites For Learning Programming
https://www.informationweek.com/cloud/software-as-a-service/10-great-websites-for-learning-programming/d/d-id/1321154

[7] 9 Places You Can Learn How to Code (for Free)
https://www.inc.com/larry-kim/9-places-you-can-learn-how-to-code-for-free.html

[8] 11 Websites To Learn To Code For Free In 2017
https://www.forbes.com/sites/laurencebradford/2016/12/08/11-websites-to-learn-to-code-for-free-in-2017/

[9] 6 Easiest Programming Languages to Learn for Beginners
https://www.makeuseof.com/tag/easiest-programming-languages-beginners/

[10] 7 Fundamental Tips To Learn Programming From Scratch
https://www.nextacademy.com/blog/7-tips-to-learn-programming-from-scratch/

[11] 15 Free Games to Level Up Your Coding Skills
https://skillcrush.com/2017/04/03/free-coding-games/

[12] 15 free games that will help you learn how to code
http://www.businessinsider.com/15-free-games-that-will-help-you-learn-how-to-code-2017-4/

[13] Programming Languages for Game Design
https://www.gamedesigning.org/career/programming-languages/

[14] Barker, L. J., McDowell, C., & Kalahar, K. (2009, March). Exploring factors that influence computer science introductory course students to persist in the major. In ACM SIGCSE Bulletin (Vol. 41, No. 1, pp. 153-157). ACM.

[15] Learning Basic Programming Concepts By Creating Games With Scratch Programming Environment by Ouahbia, Kaddaria, Darhmaouib, Elachqara, and Lahmine, Procedia - Social and Behavioral Sciences 191 (2015) 1479 – 1482.

[16] Barnes, T., Richter, H., Powell, E., Chaffin, A., & Godwin, A. (2007, June). Game2Learn: building CS1 learning games for retention. In ACM SIGCSE Bulletin (Vol. 39, No. 3, pp. 121-125).

[17] Tic-tac-toe
https://en.wikipedia.org/wiki/Tic-tac-toe

[18] Learn C++ Tutorial Point
https://www.tutorialspoint.com/cplusplus/index.htm